\documentclass[10pt,aps,twocolumn,superscriptaddress,preprintnumbers,amsmath,amssymb,nofootinbib]{revtex4-2}
\usepackage{epsfig}  
\usepackage{graphicx}
\usepackage{color}
\usepackage{float}
\usepackage{amsfonts}
\usepackage{amsmath}
\usepackage{slashed}
\usepackage{soul}
\usepackage[colorlinks,citecolor=blue,urlcolor=blue,bookmarks=false,hypertexnames=true]{hyperref}

\large

\begin{document}

\title{Spin-Flavor Oscillations of Relic Neutrinos in Primordial Magnetic Field }

\author{Ashutosh Kumar Alok}
\email{akalok@iitj.ac.in}
\affiliation{Indian Institute of Technology Jodhpur, Jodhpur 342037, India}

\author{Trambak Jyoti Chall}
\email{chall.1@iitj.ac.in}
\affiliation{Indian Institute of Technology Jodhpur, Jodhpur 342037, India}

\author{Neetu Raj Singh Chundawat}
\email{chundawat.1@iitj.ac.in}
\affiliation{Indian Institute of Technology Jodhpur, Jodhpur 342037, India}

\author{Arindam Mandal}
\email{mandal.3@iitj.ac.in}
\affiliation{Indian Institute of Technology Jodhpur, Jodhpur 342037, India}

\begin{abstract}
The neutrino magnetic moment operator clasps a tiny but non-zero value within the standard model (SM) of particle physics and rather enhanced values in various new physics models. This generation of the magnetic moment ($\mu_\nu$) is through quantum loop corrections which can exhibit spin-flavor oscillations in the presence of an external magnetic field. Also, several studies predict the existence of a primordial magnetic field  (PMF)  in the early universe, extending back to the era of Big Bang Nucleosynthesis (BBN) and before. The recent NANOGrav measurement can be considered as a strong indication of the presence of these PMFs. In this work, we consider the effect of the PMF on the flux of relic neutrinos. For Dirac neutrinos, we show that half of the active relic neutrinos can become sterile due to spin-flavor oscillations well before becoming non-relativistic owing to the expansion of the Universe and also before the timeline of the formation of galaxies and hence intergalactic fields, subject to the constraints on the combined value of $\mu_\nu$ and the cosmic magnetic field at the time of neutrino decoupling. For the upper limit of PMF allowed by the BBN, this can be true even if the experimental bounds on $\mu_{\nu}$ approaches a few times its SM value.
\end{abstract}

\maketitle

\newpage
\section{introduction}
\label{into}

Neutrinos that were generated in the aftermath of the Big Bang constitute the most prevalent particles within the cosmic landscape, ranking second only to the 3K black body electromagnetic radiation emanating from the Cosmic Microwave Background (CMB). The ubiquity of relic neutrinos dispersed throughout the universe serves as a perpetual testament to the underpinnings of the Hot Big Bang cosmological model. Emerging during the nascent stages of the Universe, these neutrinos underwent decoupling from the broader matter content within mere moments following the Big Bang event.
In accordance with the standard model (SM) of particle physics, the Cosmic Neutrino Background (C$\nu$B) is anticipated to encompass the three active neutrino flavors. These neutrinos are potentially of either Dirac or Majorana nature and at least two of the mass eigenstates must have mass $m_{\nu} \sim \mathcal{O}(\rm meV)$ and thereby being non-relativistic in nature at the present \cite{Esteban:2020cvm}.

The existence of the C$\nu$B has been indirectly confirmed, particularly at MeV energies, through precise measurements of the primordial abundances of light elements during the epoch of Big Bang Nucleosynthesis (BBN) \cite{Steigman:2012ve}. At later cosmic epochs, the presence of relic neutrinos has been substantiated through data on the anisotropies of the CMB \cite{Planck:2018vyg} and the distribution of Large Scale Structures (LSS) in the universe \cite{eBOSS:2020yzd, DES:2021wwk}. Nonetheless, direct observation of relic neutrinos remains elusive, a pursuit that would hold immense significance in the realm of fundamental physics. The inherent challenges arise from their weak interactions and low energy, making the practical detection of these neutrinos seemingly unattainable.

Despite these hurdles, various unconventional proposals have emerged to probe relic neutrinos. These proposals span different parameter spaces encompassing neutrino mass, temperature, and the Dirac or Majorana nature of neutrinos. For e.g.,  the PTOLEMY experiment endeavours to detect the C$\nu$B by capturing electron neutrinos on a 100 g Tritium target \cite{PTOLEMY:2018jst}. Another approach is grounded in the Stodolsky effect \cite{Stodolsky:1974aq, Domcke:2017aqj, Duda:2001hd}, wherein the neutrino background imparts changes to the energy levels of atomic electron spin states, analogous to the Zeeman effect. Furthermore, the coherent neutral current scattering process \cite{Shergold:2021evs, Opher:1974drq, Lewis:1979mu} and the passage of an accelerated ion beam through the target \cite{Bauer:2021uyj} offer additional avenues for detecting these elusive neutrinos.

The direct detection of relic neutrinos using these techniques presents an exceptionally challenging endeavour, warranting the considerable effort. Consequently, the likelihood is higher that these relic neutrinos will be detected via indirect methodologies, many of which hinge on extraterrestrial sources. Among these, a notable approach employs ultra-high energy cosmic rays possessing energies surpassing the Greisen-Zatsepin-Kuzmin (GZK) cut-off at $5\times10^{19}$ eV. In scenarios where neutrinos possess mass and undergo radiative decay, the C$\nu$B could be discerned through the annihilation of ultrahigh energy neutrinos with non-relativistic neutrinos from the relic sea \cite{Fargion:1997ft, Weiler:1997sh}.
Another approach involves the mechanism of atomic de-excitation, underpinned by the Pauli exclusion principle \cite{Yoshimura:2014hfa}. Moreover, the potential detection of relic neutrinos is intricately linked to gravitational clustering, the development of structures, and localized concentrations \cite{Dighe:1998es, Ringwald:2004np, Brdar:2022kpu}. These methods collectively contribute to the comprehensive pursuit of understanding and identifying the elusive relic neutrinos, augmenting our insights into the early universe's dynamics.

The significance of the weak interaction characteristics of neutrinos is evident in the majority of these techniques, underscoring the essential role that the density of active relic neutrinos can play as a critical parameter. This density stands as a determining factor, impacting the effectiveness and practicality of a range of detection methods aimed at elucidating the mysteries surrounding relic neutrinos and their fundamental implications for the early history of the universe.

Relic neutrinos transitioned from being ultra-relativistic to non-relativistic during the early stages of cosmic evolution. The time of this transition depends on the expansion rate of the universe and the temperature of the background plasma.
This transition occurred when the temperature of the universe dropped to a point where the neutrinos' thermal energies became comparable to their rest masses. These transition temperatures are, thus, characteristic to the considered neutrino masses. The chiral oscillations set in at these temperatures result in the conversion of the non-relativistic active neutrinos into opposite helicity states.

There can be a twist in the story with neutrinos possessing a finite magnetic moment ($\mu_{\nu}$) owing to their non-zero masses \cite{Giunti:2014ixa}. In the context of the Minimal Extended Standard Model (MESM), where the right-handed neutrinos in three generations constitute the sole additional gauge-singlet fields, the diagonal elements ($\alpha = \beta$) of the magnetic moment operator \(\mu_{\nu}^{\alpha\beta}\) are approximately of the order \(10^{-19}\,\mu_B\) \cite{Lee:1977tib,Fujikawa:1980yx}. However, there are several experiments which put an upper bound on the values of neutrino magnetic moments, such as the TEXONO \cite{TEXONO:2006xds}, the GEMMA reactor neutrino experiment \cite{Beda:2012zz} and the BOREXINO experiment \cite{Borexino:2017fbd} which uses the solar neutrinos as a source. The bounds obtained from these experiments are of the order of $10^{-11}\, \mu_B$. The latest upper limit is $\sim 10^{-12}\, \mu_B$ which is provided by the XENON experiment \cite{XENON:2022ltv}. All of the above-mentioned experimental limits are, however, much greater than the SM predicted value.  The value of neutrino magnetic moment can be enhanced up to the current experimental upper limit in various beyond  SM scenarios. A finite value of neutrino magnetic moment can lead to the phenomenon of spin-flavor oscillation (SFO) in the adjacency of an external magnetic field, rendering sterile neutrinos (various ramifications of neutrino magnetic moments, including SFO can be found in a number of recent works; see e.g., \cite{Brdar:2020quo,Alok:2022pdn,SinghChundawat:2022mll,Kopp:2022cug,Zhang:2023nxy,Brdar:2023tmi,Koksal:2023qch,Grohs:2023xwa,Sasaki:2023sza,Bulmus:2022gyz,MammenAbraham:2023psg,Jana:2022tsa,Carenza:2022ngg,Baym:2020riw,Giunti:2023yha,Denizli:2023rqe}). Therefore, a legitimate question to ask is whether half of the active neutrinos were converted to sterile neutrinos under the effect of the underlying magnetic field via SFO, even before the appearance of the chiral oscillations due to the expansion of the universe. This is the primary focus of this work. Two types of magnetic fields can account for such effects - the primordial magnetic field  (PMF) and the intergalactic magnetic field. However, for neutrino mass above $10^{-3}$ eV, the relic neutrinos are expected to become non-relativistic even before the formation of the galaxies. Thus, intergalactic magnetic fields become irrelevant in this context, and hence, the quest shifts to the contribution of PMF.

Following their decoupling, neutrinos persisted within the possible magnetic field environment of the early universe, a phenomenon that has received significant attention in the literature \cite{Harrison:1969,Hogan:1983zz,Baym:1995fk,Quashnock:1988vs,Grasso:1997nx,Sigl:1996dm,Subramanian:2015lua,Maleknejad:2012fw,Vachaspati:2020blt,Grasso:2000wj,Steinmetz:2023nsc,Rafelski:2023zgp,Mtchedlidze:2021bfy,Planck:2015zrl,POLARBEAR:2015ktq,Neronov:2010gir,Tavecchio:2010mk,Taylor:2011bn,Dolag:2010ni}. This magnetic field context has recently been subject to analysis leveraging observations from NANOGrav \cite{Li:2023yaj,RoperPol:2022iel,NANOGrav:2023gor,EPTA:2023fyk,Reardon:2023gzh,Xu:2023wog}. In this work, we consider the intriguing proposition of the transformation of active neutrinos to sterile ones propelled by the dynamics of SFO in the presence of pervasive PMF prevalent during the nascent epochs of the universe. Considering Dirac neutrinos, we study the possibility of depletion of active neutrino population by half, via sufficient SFO and point out the requisite parameter space of $\mu_\nu$ and PMF at the time of neutrino decoupling, subject to the bound on the same derived from primordial nucleosynthesis.  If such a process under SFO reduces the population of active neutrinos at early epochs, that can have several implications. For example, in  \cite{Green:2021gdc}, the weak interactions of such right-handed neutrinos with dark matter were studied. It was pointed out that such interaction is possible in late times since the right-handed population of neutrinos was presumed to be present only when they were in a non-relativistic state. However, if right-handed neutrinos populate significantly, then they may take part in such interactions at earlier times as well. 

The plan of this study is outlined as follows. Sec. \ref{sec:sfp} provides an elaborate exposition of SFO, encompassing the requisite conditions and the methodology employed for averaging out. Sec. \ref{pmf}  discusses the origin and bounds on PMF. Subsequently, the outcomes of our investigations are presented in the ensuing section. The concluding observations are summarized in Sec. \ref{conc}.

\section{Neutrino spin-flavor oscillations}
\label{sec:sfp}
In the present-day universe, the relic neutrinos are considered non-relativistic based on observations of the C$\nu$B temperature. However, during the early stages of the universe, when the background temperature was extremely high, these neutrinos were in an ultra-relativistic state. Since the study focuses on understanding the behaviour of relic neutrinos even before they became non-relativistic, the study employs the methodology commonly used for analyzing ultra-relativistic neutrinos.  This approach allows us to delve into the characteristics and dynamics of relic neutrinos during the early stages of the universe when they remained in an ultra-relativistic state due to the prevailing high temperatures.

In our analysis, we assume the neutrinos to be Dirac in nature.
The trajectory of the neutrino mass eigenstate $\nu_{i}^{s}$ evolves according to the Dirac equation in the presence of a magnetic field  \cite{Popov:2019nkr},
\begin{equation}
     (\gamma_{\mu}p^{\mu} - m_{i} - \mu_{i}\boldsymbol{\Sigma B} ) \nu_{i}^{s}(p) = 0, 
\end{equation}
where $s\, (= \pm 1)$ being the eigenvalues of the spin operator $\hat{S}_{i}$ and $\mu_i$ stands for the diagonal magnetic moment of neutrino. Here, we work under the assumption that the Dirac neutrinos do not possess any transition magnetic moments \cite{Popov:2019nkr,Lichkunov:2020lyf}. This can be a valid approximation, for example, in the MESM, where the magnetic moments of neutrinos are given by \cite{Giunti:2014ixa}
\begin{equation}
 \mu_{ij}\simeq \frac{3 G_{\mathrm{F}}}{16 \sqrt{ 2}  \pi^2}\left(m_i \pm m_j\right) \left(\delta_{ij}-\frac{1}{2} \sum_{l=e,\mu,\tau} U_{l i}^* U_{l j} \frac{m_l^2}{m_{W}^2}  \right).
 \label{mom}
\end{equation}
Here, $i$ and $j$ are mass indices, $U$ stands for the PMNS mixing matrix, $G_F$ is the Fermi constant, $l$ is the lepton flavor index and $m_W$ is the mass of the W gauge boson.
It is clear that the off-diagonal transition moments are highly suppressed with respect to the diagonal ones due to the presence of the ratio ${m_l^2}/{m_{W}^2}$ in eq. (\ref{mom}).

The spin operator $\hat{S}_{i}$, commutes with the Hamiltonian ($\hat{H}_{i}$) in the presence of magnetic field. $\hat{H}_{i}$ and $\hat{S}_{i}$ are given by,
\begin{equation}
    \hat{H}_{i} = \gamma_{0} \boldsymbol{\gamma p} + \mu_{i}\gamma_{0} \boldsymbol{\Sigma B} + m_{i} \gamma_{0}, 
    \label{hamiltonian}
\end{equation}
and
\begin{equation}
    \hat{S}_{i} = \frac{m_{i}}{\sqrt{m_{i}^{2}\boldsymbol{B}^{2} + \boldsymbol{p}^{2}\boldsymbol{B}_{\perp}^{2}}} 
    \left[ \boldsymbol{\Sigma B} - \frac{i}{m_i} \gamma_{0} \gamma_{5} [\boldsymbol{\Sigma \times \boldsymbol{p}}] \boldsymbol{B} \right]\,,
    \label{spin-op}
\end{equation}
respectively. Neutrinos are assumed to propagate along the positive $z$-axis, thus the momentum of neutrino is $\boldsymbol{p} = p_{z}$ and the  magnetic field is given by, $\boldsymbol{B} = (B_{\perp}, 0, B_{\parallel})$. The energy of neutrino is given as \cite{Popov:2019nkr},
\begin{equation}
    E_{i}^{s} = \sqrt{m_{i}^2 + p^2 + \mu_{i}^{2} \boldsymbol{B}^{2} + 2\mu_{i}s\sqrt{m_{i}^{2}\boldsymbol{B}^{2} + \boldsymbol{p}^{2}\boldsymbol{B}_{\perp}^{2}} }\,.
    \label{en-gen}
\end{equation} 

For ultra relativistic neutrinos $p \gg m_i$ and $p \gg \mu_\nu B$. With these approximations,
\begin{equation}
    E_{i}^{s} \approx p + \frac{m_{i}^{2}}{2p} + \frac{\mu_{i}^{2}\boldsymbol{B}^{2}}{2p} + \mu_{i}sB_{\perp}. 
    \label{energy}
\end{equation}

The probability of SFO is given as,
\begin{equation}
     P_{\alpha \beta}^{h h'}(x) = |\langle \nu_{\beta}^{h'}(0)|\nu_{\alpha}^{h}(x)\rangle|^{2}\,.
     \label{prob}
 \end{equation}
 This expression essentially quantifies the likelihood of a transformation from the initial flavor, denoted as  $\alpha$, with a specific  handedness $h$, to the eventual flavor $\beta$, also with a specific handedness, denoted as $h'$ ($h$'s denote left or right handedness). Expanding the probability expression \eqref{prob},  using the neutrino mass eigenstates ($\nu_{i}^{s}$) allows us to represent the oscillation probability as a sum over all possible mass eigenstates:
 \begin{eqnarray}
     P_{\alpha \beta}^{h h'}(x) &=& \delta_{\alpha \beta}\delta_{h h'}\nonumber \\
     &-& 4\sum_{\{i,j,s,s'\}}{\rm{Re}}([A_{\alpha\beta}^{h h'}]_{i,j,s,s'})\sin^{2}\left(   \frac{E_{i}^{s}-E_{j}^{s'}}{2}\right)x \nonumber\\
     &+&2\sum_{\{i,j,s,s'\}}{\rm{Im}}([A_{\alpha\beta}^{h h'}]_{i,j,s,s'})\sin\left( E_{i}^{s}-E_{j}^{s'}\right)x,
     \label{prob-1}
 \end{eqnarray}
 where $[A_{\alpha\beta}^{h h'}]_{i,j,s,s'} = U^{*}_{\beta i}U_{\alpha i}U_{\beta j}U^{*}_{\alpha j} (C_{is}^{h'h})(C_{js'}^{h'h})^{*}$ and $C_{is}^{h'h} = \langle\nu_{i}^{h'}|\hat{P}_{i}^{s}|\nu_{i}^{h}\rangle\,,$ through which we can get the amplitude of transition from $h$ to $h'$:
 \begin{equation}
     \langle\nu_{i}^{h'}\left( t \right)|\nu_{i}^{h}\left( 0 \right)\rangle = \sum_{s} C_{is}^{h'h} e^{-iE_{i}^{s}t}\,.
    \label{amp}
 \end{equation}  
 
 By substituting this inner product into eq.~\eqref{prob}, we obtain the  expression for the probability \eqref{prob-1}. The projection operator is defined as $\hat{P}{i}^\pm = \frac{1 \pm \hat{S}{i}}{2}$. Additionally, the summation degeneration is
 \begin{equation}
 \sum_{\{i,j,s,s'\}} = \sum_{i>j;s,s'} +\sum_{s>s';i = j}.
 \label{sum}
\end{equation}
By examining the expression \eqref{prob-1}, we observe that the phase of SFO is determined by the energy difference, denoted as $\left( E_{i}^{s}-E_{j}^{s'}\right)$. This energy difference can be derived from the previously defined expression \eqref{energy}, and when assuming that all neutrino states possess an equal magnetic moment, the phase $\phi$   can be represented as follows:
\begin{equation}
    \phi = \left[\frac{\Delta m_{ij}^{2}}{2p} + \mu_{\nu}(s-s')B_\perp\right]x.
    \label{phase}
\end{equation}

The phase mentioned in above equation can be divided into two distinct components. The initial term within eq.~ \eqref{phase} represents the phase contribution attributed to vacuum oscillations. This vacuum oscillation phase, denoted as $\phi_v$, is given by  
\begin{equation} 
\phi_v=\frac{\Delta m_{ij}^{2}}{2p}x \,.
\end{equation}
Here, $\Delta m_{ij}^{2}$ is the squared mass difference between the neutrino mass eigenstates $\nu_i$ and $\nu_j$. On the other hand, the second component of the phase arises due to the presence of an external magnetic field. This magnetic field-induced phase, denoted as $\phi_B$, is expressed as:
\begin{equation} 
\phi_B=\mu_{\nu}(s-s')B_\perp x\,.
\end{equation} 
Therefore, the total phase contributing to the SFO effect can be seen as a combination of these two distinct contributions, the vacuum oscillation phase and the magnetic field-induced phase.

Consequently, the frequency associated with this SFO induced by the magnetic field is given by
\begin{equation}
    \omega_B = \mu_{\nu}(s-s')B_\perp\,.
\end{equation}
We observe that in order to achieve an effective averaging process, a significant number of oscillation cycles, denoted as ${\mathcal N}$, is required. For a sufficiently large number of SFO, typically on the order of at least 100, the oscillation probability becomes averaged and adopts the following form:
 \begin{equation}
     P_{\alpha \beta}^{h h'} = \delta_{\alpha \beta}\delta_{h h'} - 2\sum_{\{i,j,s,s'\}}{\rm{Re}}([A_{\alpha\beta}^{h h'}]_{i,j,s,s'})\,.
     \label{prob-avg}
\end{equation}
Therefore, the probability associated with the survival of active left-handed neutrinos can be expressed as
 \begin{equation}
     P_{\alpha \beta}^{LL} = \delta_{\alpha \beta} - 2\sum_{\{i,j,s,s'\}}{\rm{Re}}([A_{\alpha\beta}^{LL}]_{i,j,s,s'})\,.
     \label{prob-avg-l}
\end{equation}
Expanding the expressions for $A$'s using the definition given in eq.~\eqref{sum}, we arrive at $P_{\alpha \beta}^{LL} = 1/2$ \cite{Alok:2022pdn}. Consequently, this implies that half of the left-handed neutrinos survive the process, while the remaining half undergoes conversion to right-handed neutrinos.

\section{Primordial magnetic field: Origin and bounds}
    \label{pmf}

Magnetic fields exhibit a pervasive presence across a wide spectrum of scales that have been investigated, ranging from planets and stars to interstellar and even intergalactic spaces. Irrespective of age or type, all galaxies seem to possess a static magnetic field with a strength of a few $\rm \mu G$ or less. Magnetic field of similar strengths are observed within galaxy clusters. Intriguingly, even the intergalactic medium within voids is suspected to contain a faint magnetic field, around $\sim 10^{-16}$ G in strength, extending coherently over Mpc scales. The existence of such a magnetic field filling the volume of void regions presents a challenge for explanation based on purely astrophysical processes and hints at the potential favorability of a primordial origin. Such a relic magnetic field can serve as a seed field for amplification through dynamo processes within collapsed objects and may account for the observed intergalactic magnetic fields. In this section, we will provide a concise overview of the origin of PMF, along with a discussion of the existing observational constraints on their strengths.

At first, the origin of the PMF was attributed to primordial vorticity, as postulated by Harrison \cite{Harrison:1969}. The Harrison mechanism posits that during the expansion of the universe, electrons and ions undergo different rates of angular velocity reduction. This disparity results in an electromotive force, which, in turn, produces electric currents. As a consequence, these electric currents give rise to the formation of a magnetic field. Over time, various alternative origins of vorticity leading to the generation of magnetic fields have been proposed. These vortices are linked to significant phase transitions that occurred during the early universe, such as the Quantum Chromodynamics (QCD) phase transition, which occurred when the universe's temperature had cooled to approximately $T \sim 200$ MeV, and the electroweak (EW) phase transition, which took place at a higher temperature of around $T \sim 100$ GeV.

During these phase transitions (PT), the universe experienced substantial alterations in its fundamental characteristics, including the symmetry-breaking of fundamental forces. These transitions were accompanied by the release of vast amounts of energy, resulting in the creation of topological defects that produced vorticity within the primordial plasma. This vorticity, in turn, gave rise to the generation of magnetic fields through diverse mechanisms, including the Biermann battery effect.

In the SM, both the Quantum Chromodynamics Phase Transition (QCDPT) and the Electroweak Phase Transition (EWPT) are not first-order phase transitions; instead, they are described as analytic crossovers, as elaborated in \cite{Subramanian:2015lua}. In the context of Supersymmetric (SUSY) extensions of the SM, such as the Minimal Supersymmetric SM, there are parameter regimes where the EWPT can exhibit characteristics of a first-order phase transition. Additionally, if the lepton chemical potential in neutrinos falls within the cosmologically permissible range, the QCDPT's nature could transition from a crossover to a first-order phase transition.

In the case of first-order phase transitions, bubbles of the new phase nucleate within the old phase, giving rise to initial seed magnetic fields. As these bubbles expand and collide, the magnetic fields can undergo further amplification through a process known as the dynamo mechanism \cite{Hogan:1983zz}. This mechanism can lead to the growth of the field strength from their initial seed values. The dynamo mechanism provides a plausible explanation for the generation of magnetic fields during both the EW \cite{Baym:1995fk,Grasso:1997nx} and QCD \cite{Quashnock:1988vs,Sigl:1996dm} phase transitions. The origin of PMF can also be linked to the inflationary epoch in the early universe, as proposed in \cite{Maleknejad:2012fw} where the initial strength of $(B \sim 10^{46}\, \rm G)$ has been shown. During this period, the breaking of the conformal invariance in the electromagnetic action leads to the generation of such a substantial magnetic field.

The strength of the PMF diminishes as the universe expands, a phenomenon characterised by the time-varying expansion parameter $a(t)$ \cite{Subramanian:2015lua,Vachaspati:2020blt,Grasso:2000wj,Steinmetz:2023nsc}. The evolution of the PMF can vary from linear to non-linear, and it can exhibit Gaussian random behaviouur. Notably, the vacuum fluctuations of the electromagnetic field during the inflationary period are amplified, giving rise to Gaussian random stochastic fluctuations in the magnetic field. However, for sub-Hubble scale fields generated during the EWPT and QCDPT, there may be some non-Gaussianity \cite{Subramanian:2015lua, Mtchedlidze:2021bfy}. These characteristics bear significant implications for our comprehension of the early universe and the processes that shaped its magnetic fields.

The definitive identification of evidence supporting a PMF would stand as a momentous discovery, introducing an entirely novel observational lens through which we can peer into the intricate processes that unfolded during the nascent stages of the Universe. Significantly, a multitude of recent observational initiatives has bestowed us with robust limitations on the potential strength of the PMF. These include bounds coming from Cosmic Microwave Background (CMB) anisotropy, CMB B-mode polarization, high energy astrophysical sources as well as  nanohertz gravitational measurements.

 PMF can influence the polarization of CMB photons through Faraday rotation, leading to observable distortions in CMB maps. Analyzing these distortions provides constraints on PMF strength and correlation length. The Planck 2015 data \cite{Planck:2015zrl} offers the most comprehensive analysis, limiting PMF strength to a few nG at present, depending on the considered model. Specifically, for nearly scale-invariant PMFs, the upper limit is $B < 2.1$ nG on Mpc scales with a 95\% confidence level.
The POLARization of the Background Radiation (POLARBEAR) experiment has contributed to constraining PMFs \cite{POLARBEAR:2015ktq}. By comparing predicted B-mode polarization (specifically from vector modes) with observed data at high multipoles, the experiment sets a limit of $B_0 <$ 3.9 nG at a 95\% C.L.

Hess and Fermi observations of TeV blazars \cite{Neronov:2010gir} have revealed the presence of intergalactic magnetic fields spanning the space between galaxies. The emission of TeV gamma-rays by blazars initiates a process where these photons interact with background starlight, generating electron-positron pairs. These pairs then undergo Compton scattering on CMB photons, converting them into GeV gamma-rays. The expected detection of GeV photons from blazars remains elusive. The most straightforward interpretation suggests that electron-positron pairs are deflected by magnetic fields, deviating from their anticipated light cone trajectory. This scenario imposes a lower limit on the magnetic field strength, with an estimated value of B $\gtrsim 3\times10^{-15}$ G, considering a coherence scale approximately on the order of kiloparsecs \cite{Tavecchio:2010mk,Taylor:2011bn}.
 Pulsar timing arrays (PTAs) like NANOGrav have played a significant role in tightly constraining the strength of the PMF. For instance, a recent study \cite{RoperPol:2022iel} demonstrated that the presence of a magnetic field can indeed lead to the production of such gravitational waves. The estimated present magnitude of the PMF strength falls within the range of approximately $(0.01 - 0.1)$ nG \cite{RoperPol:2022iel}.

\section{Results and Discussions}
\label{res}
It should be emphasized that even after neutrino decoupling, these particles retained their relativistic nature due to the exceedingly high background temperature prevailing during that period. However, as the universe expands, this background temperature decreases with the scale factor, as described by the equation:
\begin{equation}
T(t) = \frac{T_0}{a(t)},
 \label{temp}
\end{equation}
where $T_0$ stands for the temperature of the C$\nu$B at present, $T(t)$ and $a(t)$ are the temperature and scale factor at any given time $t$, respectively. The present C$\nu$B temperature, which can be indirectly derived from the temperature of the CMB, is approximately $1.95$ K. Notably, this temperature range aligns with the feasible mass spectrum of neutrino eigenstates, thus rendering neutrinos in their current non-relativistic state. The transition from relativistic to non-relativistic states for neutrino eigenstates occurs at distinct temperatures depending  upon their masses.

The observation of neutrino oscillations implies the existence of neutrino mass, although one can only determine the differences in squared masses of neutrinos  denoted as $\Delta m^{2}_{21}$ and $\Delta m^{2}_{31}$. Current global fits impose constraints on the neutrino mass-squared splittings, yielding values $\Delta m_{21}^{2}= 7.42\times 10^{-5}$\,\ eV$^{2}$ and $|\Delta m_{31}^{2}| = 2.51\times 10^{-3}$\,\ eV$^{2}$, with the sign of the latter still under exploration \cite{Esteban:2020cvm}.  Additionally, cosmological observations contribute by placing limits on the sum of neutrino mass eigenstates. An upper bound of $\sum_{i}{m_i}  < 0.26$\,\ eV was established in \cite{Loureiro:2018pdz}, in alignment with neutrino oscillation experiments, leading to an upper limit for the lightest neutrino mass species, ${m_{\nu}} < 0.086$\,\ eV. These combined constraints indicate that at least two of the mass eigenstates must have mass $m_{\nu} \sim \mathcal{O}(\rm meV)$ \cite{Esteban:2020cvm} , while the heaviest eigenstate's mass is expected to be approximately within the range of $m_{\nu} \sim 0.1 \,\ \rm eV $. In this work, we consider neutrino masses within the range $(0.001 - 0.1)$ eV. 

 For $m_{\nu_i} = 0.1$ eV,   the transition from relativisic to non-relativistic state should occur around $T\sim 770$ K when its energy becomes comparable to its mass. Similarly, for the lightest conceivable neutrino mass, this transition is expected to occur at 
$T \sim 7.7$ K, which approximately corresponds to the time of galaxy formation, roughly $\sim 10^{9}$ years after the universe's inception. Thus, up until this point, all neutrino mass eigenstates would have transitioned to a non-relativistic state.

It is imperative to note that the influence of intergalactic magnetic fields does not bear relevance within the scope of this study, which centers on the relativistic relic neutrinos. This era, identified as the ``dark era", predates the formation of the universe's first galaxies. Consequently, the study exclusively investigates the impact of PMF on neutrinos following their decoupling from the rest of the matter, specifically through the phenomenon of SFO. As briefly introduced in Sec. \ref{pmf}, the early universe could have potentially harbored a substantial magnetic field, a phenomenon that might have been initiated through various mechanisms. In this scenario, relic neutrinos existing within the influence of such a PMF would be subjected to SFO, provided the magnetic moment $\mu_{\nu}$ is of sufficient magnitude. The cumulative effect of numerous oscillations enables the probability of oscillations to average out, resulting in a value of 
$1/2$. This intriguingly leads to the conversion of half of the active neutrinos into sterile neutrinos, as elaborated in \cite{Alok:2022pdn}. To enact this conversion, we stipulate a minimum requirement of 100 cycles of SFO.

The number of oscillations completed at any given time after neutrino decoupling can be computed as,
\begin{equation}
    \mathcal{N}(t) = \int_{t_d}^{t} \frac{dt'}{T(t')}\,,
    \label{Nosc}
\end{equation}
where \(T(t')\) is the time period of SFO, which takes the form \(T = \pi/2 \mu_{\nu} B\) and is itself time-dependent due to the varying magnetic field over time. As discussed in Sec. \ref{pmf}, the PMF experiences redshift over time due to the expansion of the universe. The evolution of the PMF can be modelled by a simple scaling like \cite{Subramanian:2015lua,Vachaspati:2020blt,Grasso:2000wj,Steinmetz:2023nsc},
\begin{equation}
    B(t) \propto \frac{1}{a^{2}(t)},
    \label{evol}
\end{equation}
where \(a(t)\) is the scale factor.

However, considering the effect of physical spatial scale as the co-moving length of the magnetic field domain, the root mean squared strength of the PMF at any later moment after neutrino decoupling is given by the following power law scaling:
\begin{equation}
    B(t,L) = \left(\frac{a(t_d)}{a(t)}\right)^{2}\left(\frac{L(t_d)}{L(t)}\right)^{p}B(t_d),
    \label{scale}
\end{equation}
where $L\left(t\right)$ denotes the co-moving coherence length corresponding the magnetic field $B\left(t\right)$ and  \(t_d\) is the time corresponding to neutrino decoupling $\simeq 1$ MeV. Again, this coherence length scales as $L(t) \propto a(t)$ implying the variation of magnetic field over time as,
\begin{equation}
    B(t,L) = \left(\frac{a(t_d)}{a(t)}\right)^{2+p}B(t_d)\,.
    \label{scale-final}
\end{equation}
We can see here that we have a power index $p$ in the length scaling factor of eq. \eqref{scale}, which is an unknown parameter  that can take three values: $0.5,\,1,\, 1.5$ subject to the statistical properties of the Gaussian random magnetic field. For e.g., if the magnetic field vector is performing a random walk in 3D volume, where the number of steps can be given by $L(t)/L(t_d)$, the scaling of the field will be $\sim \left(L(t)/L(t_d)\right)^{-\frac{3}{2}}$ \cite{Hogan:1983zz}. Hence, in this case,  the index $p$ in eq. \eqref{scale} will be 1.5. Similarly, an argument based on the statistical independence of the conserved flux gives the scaling of the field as $\sim \left(L(t)/L(t_d)\right)^{-1}$\cite{Vachaspati:1991nm} (correspondingly $p=1$) and that of the field in neighbouring cells predicts the scaling as $\sim \left(L(t)/L(t_d)\right)^{-\frac{1}{2}}$\cite{Enqvist:1993np} (correspondingly $p=0.5$).

Utilizing eq. \eqref{Nosc},  the number of oscillations can be expressed as:
\begin{equation}
    \mathcal{N}(t) = \int_{t_d}^{t} \frac{2\mu_{\nu} B(t_d) a^{2+p}(t_d)}{\pi\, a'^{2+p}(t')} dt'.
    \label{Nosc-Bt}
\end{equation}

Here, \(t'\) is related to the scale factor as \(dt' = \frac{da'}{a'\,H(t')}\), and the Hubble parameter \(H\) can be written in terms of its present value as:
\begin{equation}
    H^{2}(t') = H^{2}_0 a'^{-4} ( \Omega_{R,0}+\Omega_{M,0} a'+ \Omega_{K,0} a'^{2}+\Omega_{\Lambda,0} a'^{4})\,.
    \label{Ht}
\end{equation}
Here, \(\Omega_{R,0}\), \(\Omega_{M,0}\), \(\Omega_{K,0}\), and \(\Omega_{\Lambda,0}\) signify the density parameters corresponding to radiation, matter, curvature, and dark energy, respectively and $H_0$ is the value of Hubble parameter at the present time. Observations suggest that the universe is flat, leading to the exclusion of \(\Omega_{K,0}\) from the term quadratic in \(a'\). The present-day value of Hubble parameter is $H_0 = 67.3\, \rm km\,s^{-1}\, Mpc^{-1}$ and the values of the remaining three density parameters are: \(\Omega_{R,0} = 9.24 \times 10^{-5}\), \(\Omega_{M,0} = 0.315\), and \(\Omega_{\Lambda,0} = 0.685\) \cite{Lahav:2022poa}. Using this relation, eq. \eqref{Nosc-Bt} can be written as
\begin{eqnarray}
    \mathcal{N}(t) &=& \frac{2\mu_{\nu} B(t_d) a^{2+p}(t_d)}{\pi H_0} \nonumber\\
    & \times & \int_{a(t_d)}^{a(t)} \frac{da'}{a'^{p+1}\sqrt{\Omega_{R,0}+\Omega_{M,0} a'+\Omega_{\Lambda,0} a'^{4}}}.
\end{eqnarray}
It is noteworthy that, in addition to the diminishing strength of the PMF, the number of oscillations is also influenced by the correlation between \(dt'\) and the scale factor. The value of \(\mathcal{N}\) experiences rapid growth immediately after neutrino decoupling, but the rate of this augmentation diminishes as the universe expands. Consequently, for a ``rapid" execution of a substantial number of oscillations, the PMF strength at the time of neutrino decoupling, \(B\left(t_d\right)\), must be sufficiently high for a given \(\mu_{\nu}\). If the initial field strength is relatively low, then the completion of 100 oscillations is significantly delayed due to the universe's expansion. The interpretation of the term ``quick" hinges on the specific context. In the scope of this study, our interest extends up to \(10^9\) years, as the lightest neutrino mass eigenstates maintain their relativistic nature during this period.

\begin{figure*} [htb]
\centering
\includegraphics[scale=0.9]{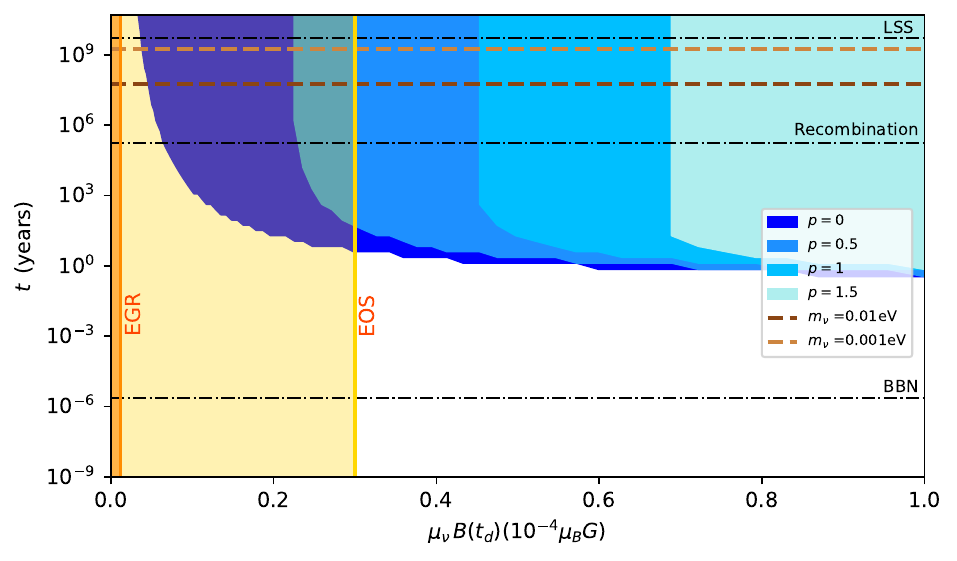}
\caption{The distinct regions for successfully averaging out the active neutrino flux via SFO for a range of the product $\mu_\nu B\left(t_d\right)$ in units of $\mu_{B}$G is depicted by the different shaded regions corresponding to the different $p$ values, where $\mu_{B}$ is Bohr magneton. EGR and EOS correspond to the upper limits on the product $\mu_\nu B\left(t_d\right)$ derived in \cite{Elmfors:1996gy} and \cite{Enqvist:1994mb,Long:2015cza}, respectively. The horizontal dashed lines correspond to the different times at which different mass species of neutrinos turn non-relativistic. The horizontal black dash-dotted lines depict the timeline for the relevant cosmological events.}
\label{muB}
\end{figure*}

In fig. \ref{muB}, we have indicated the transition times at which the relic neutrinos shift from being relativistic to non-relativistic, corresponding to different mass bounds for neutrinos as horizontal dashed lines on the time axis. This allows us to deduce whether the transition time matches or surpasses the time required for the neutrino flux to average out (i.e., the completion of 100 SFO cycles). If the averaging out time is shorter than the transition time, it implies that the total density flux of relic neutrinos is halved even before they become non-relativistic.

During the primordial nucleosynthesis, apart from adding on to the magnetic energy density as a uniform perturbation, the PMF also perturbs the $e^{+}e^{-}$ density of states, thus boosting the energy density and pressure due to pairs and also altering the $n\leftrightarrow p$ conversion rates. For the concerned ranges of PMF, its effect on the magnetic energy density acts as the dominating perturbation. So by this assumption of the dominant perturbation, an upper bound on the strength of the PMF at the very beginning of the BBN epoch  or at the time of neutrino decoupling ($T\sim 1 \rm MeV$) is obtained to be $B\left(t_d\right)\lesssim 10^{13}$ G
\cite{Kernan:1995bz,Cheng:1996yi,Kawasaki:2012va,Yeh:2022heq}. This bound is independent of whether or not neutrinos posses magnetic moment. Throughout our analysis, we do not evade this upper bound on the PMF strength.

If neutrinos have a magnetic moment, then the amount of spin oscillations of neutrinos in the early universe in the presence of PMF can be constrained by the BBN measurements. Strictly speaking, this imposes bounds on the product of neutrino magnetic moment and the strength of PMF.
Several analysis in the literature explores the possibility of the spin-flipping processes in the presence of PMF in the earlier epochs before BBN \cite{Lynn:1980gh,Fukugita:1987ag,Elmfors:1996gy,Enqvist:1992di,Semikoz:2017fbj,Enqvist:1994mb}. The interactions before BBN can not be arbitrarily frequent but rather are constrained, thanks to observational estimation of the Hubble parameter $H$. This quantity is intrinsically related to the number of generations of neutrinos ($N_{\rm eff}$). Albeit there is a discrepancy between the early and late time determination of $H$ (commonly known as Hubble's tension), it only allows a very small deviation of $N_{\rm eff}$ from its standard value  3. Any interaction resulting in a larger deviation than this can, then, be constrained. The spin-flip of neutrinos under PMF in the early universe is one such process. The neutrinos of flipped helicities, i.e., the right-handed neutrinos, add up to the total number of degrees of freedom, thereby increasing the rate of expansion of the universe. The thermal equilibrium of the interaction of neutrinos with the protons and neutrons before the onset of the BBN era fixes the ratio of light element abundances. Hence a depletion in the active neutrino population may draw serious consequences to the outcomes of primordial nucleosynthesis. In what follows, we discuss the bounds obtained on PMF and magnetic moment of the neutrinos. 

The wholesome explanation was introduced by Fukugita \textit{et al.}\cite{Fukugita:1987ag} that the BBN is affected by the neutrino spin-oscillations only if such oscillations take place at a temperature that is in between the QCDPT ($\sim 200$ MeV) and the neutrino decoupling ($\sim 1$ MeV) and some essential conditions on the spin-oscillation frequency $\left(2\mu_{\nu} B\right)$ are met. Firstly, the rate of the conversion from left to right-handed neutrinos has to exceed the expansion rate of the universe (i.e., $\Gamma_{L \to R} \gtrsim H\left(t\right)$) to bring the right-handed neutrinos into thermal equilibrium. Again, it also has to be larger than the neutrino scattering rate with the primordial plasma particles. Furthermore, the oscillation frequency has to be greater than the neutrino energy difference accounted for the distinct LH and RH neutrino refractive indices ($n_L - n_R \neq 0$).
The following limit on the product \(\mu_{\nu}B\) at present time is derived under these assumptions in \cite{Fukugita:1987ag}:
\begin{equation}
    \mu_{\nu}B\left( T_{0}\right)\lesssim 10^{-25}\,\ \mu_{B}G\,.
    \label{Fuku}
\end{equation}
A refined treatment from Elmfors \textit{et al.}\cite{Elmfors:1996gy} inculcated the effect of PMF on the refractive properties of neutrinos by studying the average evolution of the global spin-polarization vector of the entire ensemble of momentum modes and estimated the average rate of depolarization $\Gamma_{\text {dep}}$ of the active LH (anti)neutrino population and compared it with the Hubble expansion parameter to put a bound on the product of \(\mu_{\nu}\) and \(B\left(t_d\right)\) by imposing the condition $\Gamma_{\text {dep}} \gtrsim H$ for which the RH population reaches thermal equilibrium at a temperature $T$.

At the time of neutrino decoupling (i.e. at $T=T_{t_d}\left(\approx 1\,\rm MeV\right)$) the upper limit on the $\mu_\nu B$ is obtained to be:
\begin{equation}
    \mu_{\nu} B(T_{t_d}) \lesssim 1.2 \times 10^{-6}\,\ \mu_{\mathrm{B}} \mathrm{G}\,.
    \label{tight}
\end{equation}
This bound can be translated to the present and can be written as \cite{Grasso:2000wj}:
\begin{equation}
    \mu_{\nu} B(T_{0}) \lesssim 7 \times 10^{-26}\,\ \mu_{\mathrm{B}} \mathrm{G},
    \label{tightpres}
\end{equation}
which only slightly differs from eq.~\eqref{Fuku}. In the rest of the analysis we refer to this limit as EGR.

In the case where right-handed neutrinos decouple before QCDPT, the bulk of entropy released at the phase transition can suppress the relative abundances of sterile RH neutrino population to the acceptable levels, where their presence would not significantly affect the outcomes of BBN.
Thus, substantial restrictions can be imposed on the product of the neutrino magnetic moment and magnetic field strength from the measurements at BBN, which requires the thermal production of RH neutrinos to stop by the QCDPT as explained by Enqvist \textit{et al.} \cite{Enqvist:1992di} where a collision-less treatment of LH neutrinos was considered and in the chiral conversion rate, the potential $V$ was shown to account for the thermal background at $T\ll M_{W}$, where $M_{W}$ stands for the mass of the W-boson. They obtained the following bound on the product at the QCD epoch considering a homogeneous and constant magnetic field:
\begin{equation}
    \mu_{\nu}B\left( T_{0}\right)\lesssim 6.5\times 10^{-34}\,\ \mu_{B}G\,.
    \label{Enq92}
\end{equation}
 Further, Enqvist \textit{et al.} \cite{Enqvist:1994mb} considered the elastic $\nu-e$ collisions in their calculation and provided the bound at the $T_{\mathrm{QCD}}$ for the case of large-scale random magnetic field and the same is translated to the present time ($T_{0}$) and given by \cite{Long:2015cza}:
\begin{equation}
    \mu_{\nu}B\left( T_{0}\right)\lesssim 3\times 10^{-30}\mu_{B}\left( \frac{L_{0}}{10 \,\rm Mpc}\right)^{-\frac{1}{2}}G\,,
    \label{TV}
\end{equation}
where $L_{0}$ represents the magnetic field domain size at present. We refer to this limit as EOS hereon.

 With sufficient strength of PMF and neutrino magnetic moment values,  half of the active neutrino population may transform into sterile neutrinos during the early stages of the universe, even preceding the chiral oscillations that naturally transpire for non-relativistic neutrinos. In fig. \ref{muB}, the required values of the product $\mu_\nu B$ at the time of decoupling for such criteria to be satisfied are shown for different values of $p$. Since this parameter ($p$) controls the evolution of the PMF alongside the simple scaling due to the expansion of the universe, the parameter space depends heavily on $p$, which is apparent from the shift in the different coloured shades characteristic to different values of $p$  in fig. \ref{muB}. As the value of $p$ increases, the dependence of the completion time of 100 oscillations on the value of PMF at the time of decoupling becomes stronger. This can be inferred from the figure that the parameter space becomes steeper as $p$ increases. This is due to the fact that higher $p$ values lead to faster decay of PMF, and given such a situation, only a huge PMF at the time of decoupling can provide the scope for a sufficiently large number of oscillations. For completeness, we also provide the parameter space for $p=0$, which corresponds to the ratification of the PMF strength only due to the expansion of the universe. Neutrinos possessing magnetic moment propagating through the underlying PMF rendering compulsatory $\mu_\nu B({t_d})$ values result in the reduction of active neutrino population by 50\%. As can be seen from the horizontal dashed lines for different neutrino masses, the process of averaging out can transpire even before the neutrinos transit into non-relativistic states. 

The vertical lines show upper limits on $\mu_\nu B({t_d})$ from different analyses under different theoretical assumptions. As the EGR limit on the product $\mu_\nu B({t_d})$ was obtained at the time of decoupling of neutrinos as given in eq.~\eqref{tight}, it has been put in directly. Conversely, other such bounds are obtained at the time of $T_{\rm QCD}$, which were further translated to the present-day upper limit. In such cases, we have translated back the limits to the time of decoupling, following the same evolution taken to render the present-day upper limits in the corresponding analysis. The EOS limit then translates to $\mu_\nu B \lesssim 3 \times 10^{-5}\,\ \rm \mu_B \rm G$ at the time of decoupling. The EGR upper limit is shown by a bright yellow vertical line, where the shaded region denotes the allowed range of the product $\mu_\nu B$. The EOS upper limit given in eq.~\eqref{TV} is illustrated as an orange vertical line, while the shaded region on the left to this line stands for the allowed range of $\mu_\nu B$ at the time of decoupling.
\\
As evident from fig. \ref{muB}, the required strength of the product $\mu_\nu  B({t_d})$ for the depletion of left-handed neutrino population by half via SFO is forbidden by the EGR limits. However, some region of the required values of $\mu_\nu  B({t_d})$ overlap with the allowed values as obtained by the EOS limit. For $p=0$, a larger region is allowed as compared to that of higher $p$ values, which is expected, as explained above. 
For $p=0$ and $p=0.5$, some part of the required region of the product shown by different shades is allowed by the EOS limit, whereas the EGR limit forbids the entire shaded region.
On the other hand, for $p=1$ and $p=1.5$, the required values of the product to complete large enough cycles of SFO to turn half of the active neutrino population right-handed, is forbidden by both the EGR and EOS limits.

\begin{figure*} [htb]
\centering
\includegraphics[scale=0.9]{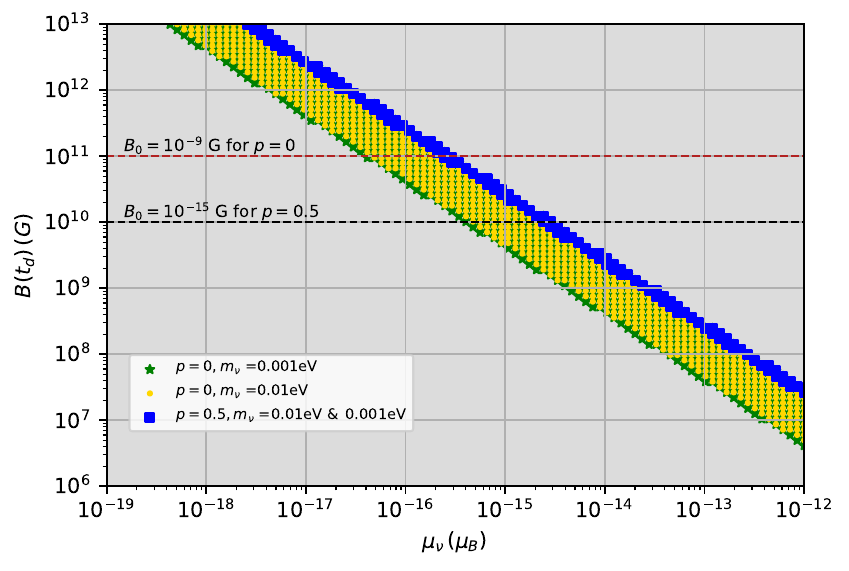}
\caption{The parameter space of $\mu_\nu$ and $B\left(t_d\right)$ corresponding to the overlapping regions of allowed and required values of $\mu_{\nu}B(t_d)$ in fig. \ref{muB}. The different colours stand for different neutrino masses and $p$ values. The red and black horizontal dashed lines depict $B(t_d)$ corresponding to the current upper and lower limit of the PMF strength at the present epoch, respectively.}
\label{muvsB}
\end{figure*}
The required values of the product can be generated by a range of values of $\mu_\nu$ and $B\left(t_d\right)$. The parameter space of $\mu_\nu$ and $B\left(t_d\right)$ rendering required values for the overlapping region in fig. \ref{muB} which allow for the averaging-out of the neutrino flux is shown in fig. \ref{muvsB}. For $p=0$, the green points denote the parameter space corresponding to the mass $m_\nu = 0.001$ eV and yellow ones represent the same for the same for $m_\nu = 0.01$ eV. Since $\mu_\nu B$ gets steeper with increasing values of $p$, the parameter space for $m_\nu = 0.001$ and $m_\nu = 0.01$ is identical for $p=0.5$, which is shown in the blue region. Thus, the averaging out may be possible even for much lower values of neutrino magnetic moment as compared to the current experimental upper limit, as per the EOS bound. For $p = 0$, a $\mu_\nu$ as small as a few times $  10^{-18}\,\ \mu_B$ would fulfil the condition for averaging out provided a sufficiently large magnetic field ($B(t_d) \sim 10^{13}\,\ \rm G$) was present at the time of decoupling. On the other hand for $p=0.5$, the required values of neutrino magnetic moment increases by one order of magnitude ($\mu_\nu \sim {\rm few} \times 10^{-17}\,\ \mu_B$), for the same requisite magnetic field. 

As suggested by various PMF generation models mentioned in Sec. \ref{pmf}, such huge magnetic fields can indeed be present at the time of neutrino decoupling, which may result in a reduction of the active neutrino population by half even before they become non-relativistic. Among those models, in \cite{Maleknejad:2012fw}, where the generation is assumed during the inflationary epoch, a magnetic field strength of $B\left(t_d\right) \sim 10^{12}$ G can be present at the time of decoupling. Also, in the post-inflationary epoch at the time of EW phase transition, the initial PMF strength mentioned in \cite{Grasso:1997nx} leads to \(B\left(t_d\right) \sim 10^{10}\) G.

Further we can also see from fig. \ref{muvsB} that for the current lower limit on the strength of the PMF at the present-day ($B_0 \sim 10^{-15}$ G), this reduction of active neutrino flux is possible only for $p=0.5$, in the allowed parameter space, where the corresponding required range on the neutrino magnetic moment is around a few times $10^{-15}\mu_{B}$. However, for the PMF evolution following the simplistic variation with $T^{2}$ i.e. the $p=0$ case, the reduction is feasible only if the present strength of the PMF is $\gtrsim 10^{-13}$G, where the corresponding requisite $\mu_\nu$ values are nearly of the order of the present upper limit of $\mu_\nu$.
For the discussed observational bounds in Sec. \ref{pmf} which are in the nG range (at the present epoch), the required $\mu_\nu$ values for the successful flux reduction as apparent from fig. \ref{muvsB} is $\sim 10^{-16}\mu_{B}$ for $p=0$. However for the $p=0.5$ case, the PMF strength at the time of decoupling turns out to be extremely large which exceeds the permitted strength provided by the BBN constraints. 

Throughout our analysis, we have considered only diagonal magnetic moments for Dirac neutrinos. In the case of Majorana neutrinos, the magnetic moment matrix has only off-diagonal elements contributing to its total magnetic dipole moment owing to its anti-symmetric nature.
In the case of two-flavor Dirac and Majorana neutrino SFO in the interstellar medium, the effect of transition magnetic moments was indeed taken into account in \cite{Kurashvili:2017zab} for ultra-high energy cosmic neutrinos. Albeit, for the three-flavor mixing case, it can be anticipated that a strong PMF can lead to these neutrino-antineutrino oscillations even for Majorana neutrinos in the early universe, which would, however, require a more comprehensive analysis for neutrinos with non-diagonal magnetic moments \cite{Lichkunov:2020lyf}, which can be studied in a future work.

\section{Conclusions}
\label{conc}
Relic neutrinos carry information about the very early stages of the universe, even before the decoupling of the photons. The fact that these relic neutrinos are non-relativistic, they undergo chiral oscillations, making half of them right-handed. If neutrinos possess magnetic moment, then such reduction of active neutrinos by half can occur through SFO under the influence of primordial cosmic magnetic field even before they become non-relativistic and also before the timeline of the formation of galaxies. This may have several implications such as impacting processes involving interaction of right-handed neutrinos in the early universe. We study such possibility of equipartition of handedness of neutrino population respecting the existing bounds on the neutrino magnetic moment as well as on the primordial magnetic field and also on their product at the time of decoupling coming from the BBN. The bounds on the product are calculated under various assumptions related to the dynamics of neutrino spin-flipping in the ambient environment before decoupling. We find that with a sufficiently large magnetic field at the time of neutrino decoupling, the reduction of active neutrino population by half is possible even for a neutrino magnetic moment greater than a few orders of magnitude above the SM value. This is possible for bounds on $\mu_\nu B(t_d)$ as obtained in \cite{Enqvist:1994mb,Long:2015cza} which we refer to as the EOS limit. Considering various evolutions of the PMF, we also obtain the $(\mu_\nu -B(t_d))$ parameter space for which the condition of averaging-out is satisfied. Within this allowed parameter space, the flux reduction is feasible even for the lower limit of the PMF at the present epoch, which is $\sim 10^{-15}$G.
Therefore, the confirmation of such a reduction in relic neutrino flux before their transition to non-relativistic states would not only provide evidence for the existence of neutrino magnetic moment but would also confirm the presence of PMF with sufficiently large strength in the nascent epochs of the universe.

\section*{Acknowledgements}
We would like to thank Joachim Kopp for engaging in valuable discussions and offering insightful suggestions on various aspects of this work. N. R. S. Chundawat extends appreciation to the organizers of the Forward Physics Facility Theory Workshop 2023 and the Department of Theoretical Physics CERN (CERN-TH) where fruitful discussions relevant to this work took place.

\end{document}